\documentclass{article}

\usepackage{arxiv}
\pdfoutput=1

\usepackage[utf8]{inputenc} 
\usepackage[T1]{fontenc}    
\usepackage{hyperref}       
\usepackage{url}            
\usepackage{booktabs}       
\usepackage{amsfonts}       
\usepackage{amssymb}       
\usepackage{nicefrac}       
\usepackage{microtype}      
\usepackage{cleveref}       
\usepackage{lipsum}         
\usepackage{graphicx}
\usepackage[numbers]{natbib}
\usepackage{doi}

\title{Surface plasmon-polaritons in graphene, embedded into medium with
		gain and losses}

\date{}

\newif\ifuniqueAffiliation

\ifuniqueAffiliation 
\author{ \href{https://orcid.org/0000-0000-0000-0000}{\includegraphics[scale=0.06]{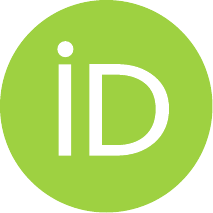}\hspace{1mm}Yu.~V.~Bludov}\thanks{Use footnote for providing further
		information about author (webpage, alternative
		address)---\emph{not} for acknowledging funding agencies.} \\
	Department of Physics, Center of Physics, and QuantaLab,\\
	University of Minho, Campus of Gualtar,\\
	4710-057, Braga, Portugal \\
	\texttt{hippo@cs.cranberry-lemon.edu} \\
	\And
	\href{https://orcid.org/0000-0000-0000-0000}{\includegraphics[scale=0.06]{orcid.pdf}\hspace{1mm}Elias D.~Striatum} \\
	Department of Electrical Engineering\\
	Mount-Sheikh University\\
	Santa Narimana, Levand \\
	\texttt{stariate@ee.mount-sheikh.edu} \\
}
\else
\usepackage{authblk}

\newbox{\orcid}\sbox{\orcid}{\includegraphics[scale=0.06]{orcid.pdf}} 
\author[1]{%
	O.~A.~Zhernovnykova%
}
\author[2]{%
	O.~V.~Popova%
}
\author[2]{%
	G.~V.~Deynychenko%
}
\author[1]{%
	T.~I.~Deynichenko%
}
\author[3]{%
	\href{https://orcid.org/0000-0001-9648-1459}{\usebox{\orcid}\hspace{1mm}Yu.~V.~Bludov\thanks{\texttt{bludov@fisica.uminho.pt}}}%
}
\affil[1]{Department of Mathematics, H.S. Skovoroda Kharkiv National  Pedagogical University, Alchevskyh Str., 29, Kharkiv, 61002, Ukraine}
\affil[2]{Department of General Pedagogy and Higher School Pedagogy, H.S. Skovoroda Kharkiv National  Pedagogical University, Alchevskyh Str., 29, Kharkiv, 61002, Ukraine}
\affil[3]{Department of Physics, Centre of Physics, and QuantaLab, University
	of Minho, Campus of Gualtar, 4710-057, Braga, Portugal}
\fi

\renewcommand{\shorttitle}{Surface plasmon-polaritons in graphene}

\hypersetup{
	pdftitle={Surface plasmon-polaritons in graphene, embedded into medium with gain and losses},
	pdfauthor={O.~A.~Zhernovnykova, O.~V.~Popova, G.~V.~Deynychenko, T.~I.~Deynichenko, Yu.~V.~Bludov},
}

\begin{document}
	\maketitle
	
\begin{abstract}
The paper deals with the theoretical consideration of surface plasmon-polaritons
in the graphene monolayer, embedded into dielectric with spatially
separated gain and losses. It is demonstrated, that presence of gain
and losses in the system leads to the formation of additional mode
of graphene surface plasmon-polaritons, which does not have its counterpart
in the conservative system. When the gain and losses are mutually
balanced, the position of exceptional point -- transition point between
unbroken and broken $\mathcal{PT}$-symmetry -- can be effectively
tuned by graphene's doping. In the case of unbalanced gain and losses
the spectrum of surface plasmon-polaritons contains spectral singularity, whose frequency is also adjustable through the electrostatic gating
of graphene.
\end{abstract}

\keywords{graphene \and surface plasmon-polariton \and $\mathcal{PT}$-symmetry}

\section{Introduction}

The physical systems, which involve both media with gain and media
with losses exhibits one specific and counter-intuitive property: being
in general non-Hermitian systems, under certain relation between gain
and losses they can possess real spectrum (similar to Hermitian system).
In the particular situation of $\mathcal{PT}$-symmetry \cite{pt-Bender1998-prl}
the gain and losses are perfectly balanced and the coordinate-dependent
complex external potential $V\left(\mathbf{r}\right)$ is characterized
by the property $V\left(\mathbf{r}\right)=V^{*}\left(-\mathbf{r}\right)$
{[}here star stands for the complex conjugation{]}. In the $\mathcal{PT}$-symmetric
structure the spectrum is real (this situation is called unbroken
$\mathcal{PT}$-symmetry) for gain/loss values below the certain threshold
and is complex for gain/loss values above this threshold. The later
situation is referred as broken $\mathcal{PT}$-symmetry and is characterized
by the presence of two modes: one is growing and another is decaying.
Although initially the conception of $\mathcal{PT}$-symmetry was
introduced in the quantum mechanical formalism, nowadays it is unclear,
whether some real quantum object described by the $\mathcal{PT}$-symmetric
Hamiltonian can be found in nature. Nevertheless, experimentally the
existence of $\mathcal{PT}$-symmetry was demonstrated in a variety
of other fields, namely mechanical systems \cite{pt-mech-Bender2013-ajp},
acoustics \cite{pt-UniVis-exp-Fleury2015-natcomm}, LRC circuits\cite{pt-lrc-exp-Schindler2011-pra},
coupled optical waveguides \cite{pt-exp-Guo2009-prl,pt-exp-Ruter2010-nature,pt-exp-Regensburger2012-nature},
or whispering-gallery resonators \cite{pt-wg-exp-Peng2014-natphys}.

Certain similarity between Maxwell and Schr\"{o}dinger equation leads to
the possibility of the realization of $\mathcal{PT}$-symmetry in
optical systems, where spatial distribution of dielectric permittivity
obeys the relation $\varepsilon\left(\mathbf{r}\right)=\varepsilon^{*}\left(-\mathbf{r}\right)$.
At the same time $\mathcal{PT}$-symmetric optical systems exhibit
a series of unusual properties like nonreciprocal (nonsymmetrical)
wave propagation \cite{pt-exp-Ruter2010-nature}, negative refraction
\cite{pt-NegRefr-Fleury2014-prl,pt-NegRefr-Valagiannopoulos2016-jopt,pt-NegRefr-Monticone2016-prx},
simultaneous lasing and coherent perfect absorption\cite{pt-PerfAbs-Longhi2010-pra,pt-PerfAbs-exp-Sun2014-prl,pt-PerfAbs-Chong2011-prl},
unidirectional visibility \cite{pt-UniVis-Mostafazadeh2013-pra,pt-UniVis-exp-Lin2011-prl,pt-UniVis-Ge2012-pra}.

Optical systems, which operation principle is based on bulk electromagnetic
waves, possess certain limit in the miniaturization of their components,
called diffraction limit. One of possible ways to overcome this diffraction
limit is to build photonic systems, which operate on surface waves
(namely, surface plasmon-polaritons) instead of bulk ones. Nevertheless,
surface plasmon-polaritons in noble metals has relatively short lifetime
due to high losses. In connection with this, using the $\mathcal{PT}$-symmetry
in plasmonics\cite{pt-plas-Alaeian2013-pra,pt-plas-Barton2018-book,pt-plas-Yang2015-scirep,pt-spp-Benisty2011-oe,pt-spp-Baum2015-jap,pt-plas-Alaeian2014-prb,pt-plas-Wang2017-prl}
can be very promising, because it could compensate losses in the noble
metals and originate lossless propagation of surface plasmon-polaritons.
Another possibility to reduce losses in plasmonics is to use a two-dimensional
carbon material graphene. Surface plasmon-polaritons sustained by
the graphene exhibit both longer lifetime and degree of localization
\cite{spp-gr-Nikitin2011-prb,spp-gr-rev-Koppens2011-nl}, if compared
to surface plasmon-polaritons in noble metals. At the same time, graphene's
conductivity can be dynamically varied through the electrostatic gating\cite{gr-cond-exp-Li2008-natp},
last fact allows to tune dynamically the wavelength \cite{spp-gr-tunable-exp-Ju2011-nn,spp-gr-tunable-Yao2014-nl,spp-gr-tunable-exp-Fei2012-nat,spp-gr-tunable-exp-Chen2012-nat,spp-gr-tunable-exp-Alonso2014-nat}
of graphene surface plasmon-polaritons (GSPPs) as well as realize tunable sensor \cite{spp-gr-sensor-tunable-Hu2016-ncom} or plasmonic modulator \cite{spp-gr-modul-Ansell2015-natcomm,spp-gr-modul-Ding2017-nanoscale}. Along with this, gated graphene
embedded into the $\mathcal{PT}$-symmetric structures allows to achieve
dynamical tunability of losses\cite{pt-graphene-tun-loss-Chatzidimitrou2018-josab}.
At the same time an optical pumping of the graphene allows the realization
of gain \cite{gr-pumping-Ryzhii2007-jap,gr-pumping-Satou2008-prb,gr-pumping-exp-BoubangaTombet2012-prb}
and, as a consequence, amplification of GSPPs\cite{spp-gr-pumping-Dubinov2011-jpcm,spp-gr-pumping-Watanabe2013-njp}.
Pumped graphene, being implemented into the lossy medium, allows
the realization of the $\mathcal{PT}$-symmetry for a series of purposes
like sensing \cite{pt-graphene-sensing-Chen2016-prappl}, waveguiding
\cite{pt-graphene-waveguide-Zhang2016-ol,pt-graphene-waveguide-Ke2018-oqe},
or diffraction grating \cite{pt-graphene-grating-Zhang2017-scirep}.

Nevertheless, the $\mathcal{PT}$-{}-symmetric structures are not
unique systems with gain or losses, which are characterized by real
spectrum. For example, a finite slab of optical gainy medium at certain
discrete frequencies possesses real spectrum\cite{near-pt-Mstafazadeh2011-pra}.
These frequencies, called spectral singularities, behave like a zero-width
resonances. Also a special relation between unbalanced gain and loss
can give rise to the generalized $\mathcal{PT}$-symmetry\cite{near-pt-generalized-Sakhdari2018-prappl},
which exhibits the same properties as its $\mathcal{PT}$-symmetric
counterpart. Along with this, systems with unbalanced gain and losses
can exhibit a series of properties like perfect absorption \cite{near-pt-absorber-Huang2016-oe},
directional coupling\cite{near-pt-Walasik2017-njp} and lossless
waveguiding\cite{near-pt-Turitsyna2017-prb}.

In this paper we consider GSPPs in graphene monolayer, cladded between
two dielectric layers: one is with gain, another with loss. We show,
that when lossless graphene is imbedded into $\mathcal{PT}$-symmetric
dielectric surrounding, the positions of the exceptional points in
GSPP can be effectively tuned by changing graphene's Fermi energy.
At the same time tunability of graphene's Fermi energy allows to vary
the positions of spectral singularities in the GSPP spectrum of lossy
graphene inside dielectric surrounding with unbalanced gain and losses.

\section{Single layer graphene in the gain-loss surrounding}

\begin{figure}
	\centering
\includegraphics[width=10cm]{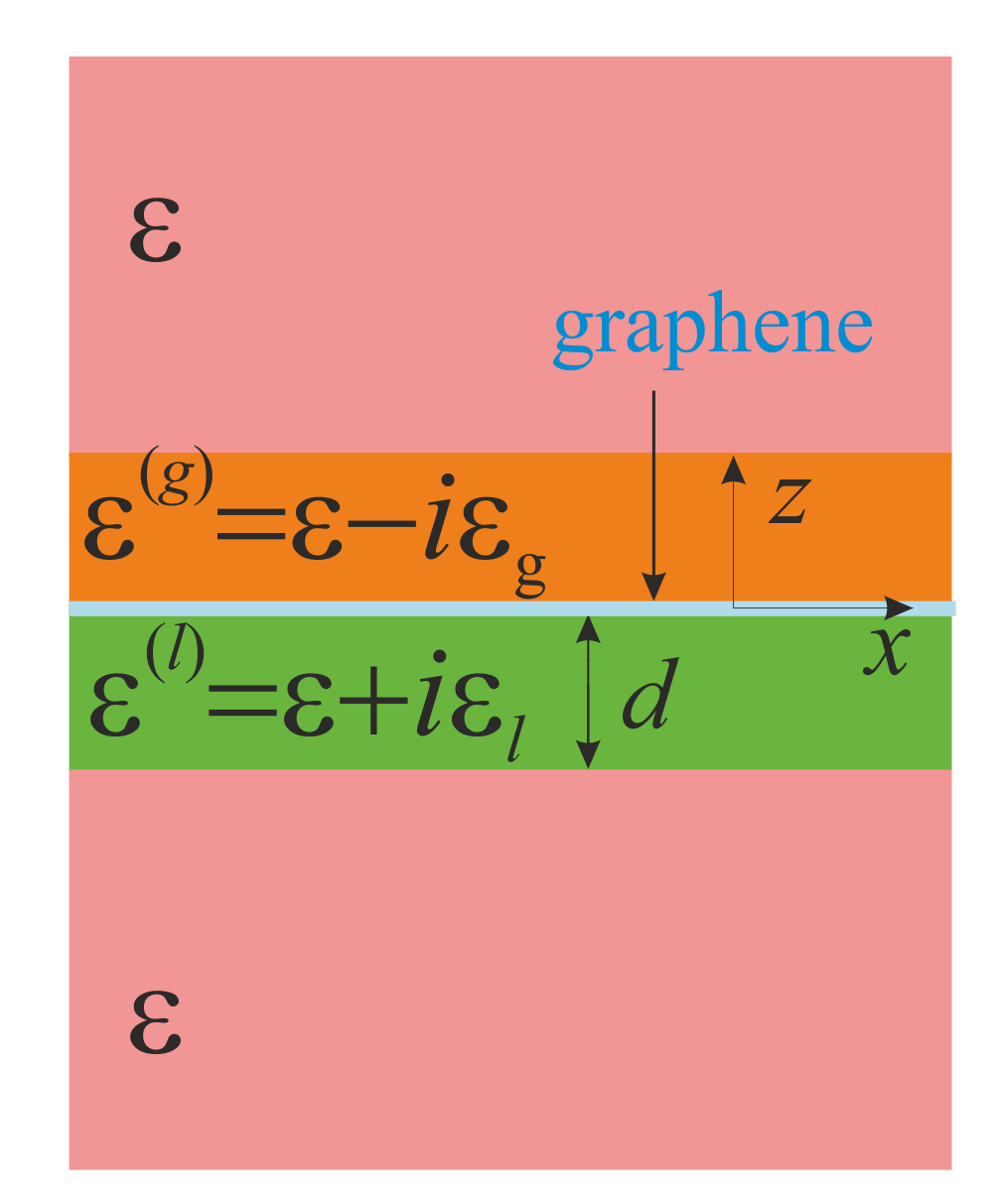}
\caption{The graphene layer, embedded into the dielectric with spatially separated gain and losses.}
\label{fig:geometry-problem}
\end{figure}
We consider the graphene layer {[}see Fig.\,\ref{fig:geometry-problem}{]}
cladded between the two dielectric media of equal thickness $d$,
one of which is arranged at spatial domain $-d<z<0$ and is characterized
by losses (dielectric constant $\varepsilon^{(l)}=\varepsilon+i\varepsilon_{l}$),
while another one occupies spatial domain $0<z<d$ and is characterized
by gain, $\varepsilon^{(g)}=\varepsilon-i\varepsilon_{g}$. The half-spaces
outside the described domains are filled with the lossless dielectric
with permittivity $\varepsilon$. 

Since GSPPs are p-polarized waves, in this paper we restrict our consideration
to the case of TM polarization, which is described by Maxwell equations
\begin{eqnarray}
-\frac{\partial H_{y}}{\partial z}=-\frac{i\omega}{c}\varepsilon\left(z\right)E_{x}+\frac{4\pi}{c}\sigma\left(\omega\right)E_{x}\delta\left(z\right),\label{eq:Max-Ex}\\
ik_{x}H_{y}=-\frac{i\omega}{c}\varepsilon\left(z\right)E_{z},\label{eq:Max-Ez}\\
\frac{\partial E_{x}}{\partial z}-ik_{x}E_{z}=\frac{i\omega}{c}H_{y}.\label{eq:Max-Hy}
\end{eqnarray}
Here we admitted that the electric field of the p-polarized wave $\mathbf{E}=\left(E_{x},0,E_{z}\right)$
possesses nonzero $x$- and $z$-components, while in its magnetic
field only $y$-component is nonzero, i.e. $\mathbf{H}=\left(0,H_{y},0\right)$.
Also in Maxwell equations (\ref{eq:Max-Ex})--(\ref{eq:Max-Hy})
we take into account the uniformity of the structure in $y$-direction
(i.e. $\partial/\partial y\equiv0$), spatiotemporal dependence of
the electromagnetic field $\sim\exp\left(ik_{x}x-i\omega t\right)$
{[}where $\omega$ is the cyclic frequency, $k_{x}$ is the in-plane
component of the wavevector, and $c$ stands for the light velocity
in vacuum{]} as well as the spatial dependence of the dielectric constant
\begin{eqnarray}
\varepsilon\left(z\right)=\left\{ \begin{array}{cc}
\varepsilon, & \left|z\right|>d\\
\varepsilon^{(l)}, & -d<z<0\\
\varepsilon^{(g)}, & 0<z<d.
\end{array}
\right.
\label{eq:varepsilon-z}
\end{eqnarray}
At the same time, in Eqs.\,(\ref{eq:Max-Ex})--(\ref{eq:Max-Hy})
Dirac delta expresses the two-dimensional character of graphene's
conductivity $\sigma\left(\omega\right)$, which can be expressed
in Drude form as
\begin{eqnarray}
\sigma\left(\omega\right)=\frac{e^{2}}{4\hbar}\frac{4E_{F}}{\pi\hbar\left(\gamma-i\omega\right)}.\label{eq:drude-cond}
\end{eqnarray}
Here $E_{F}$ is the Fermi energy and $\gamma$ is the disorder. In practice, graphene monolayer is characterized by the finite thickness $\approx 3\,$\AA. But further in the paper the thickness of graphene is supposed to be much less than thicknesses of gainy and lossy dielectric layers, so the graphene is considered as two-dimensional conductor.

In the semiinfinite spatial domain $z<-d$, the solution of Maxwell
equations (\ref{eq:Max-Ex})--(\ref{eq:Max-Hy}) can be represented
as
\begin{eqnarray}
H_{y}^{\left(-\right)}\left(z\right)=H_{y}^{\left(-\right)}\left(-d\right)\exp\left[p\left(z+d\right)\right],\label{eq:Hy-minus}\\
E_{x}^{\left(-\right)}\left(z\right)=\frac{cp}{i\omega\varepsilon}H_{y}^{\left(-\right)}\left(-d\right)\exp\left[p\left(z+d\right)\right],\label{eq:Ex-minus}\\
E_{z}^{\left(-\right)}\left(z\right)=-\frac{ck_{x}}{\omega\varepsilon}H_{y}^{\left(-\right)}\left(-d\right)\exp\left[p\left(z+d\right)\right].\label{eq:Ez-minus}
\end{eqnarray}
Here $H_{y}^{\left(-\right)}\left(-d\right)$ and $p=\left[k_{x}^{2}-\left(\omega/c\right)^{2}\varepsilon\right]^{1/2}$
are the value of magnetic field at $z=-d$ and its decaying factor,
respectively. The electromagnetic wave, whose fields are described
by Eqs.\,(\ref{eq:Hy-minus})--(\ref{eq:Ez-minus}) can be either
evanescent (when $k_{x}>\omega\sqrt{\varepsilon}/c$, and $p$ is
purely real value, which sign is chosen to be positive in this case,
$\mathrm{Re}\left[p\right]>0$), or propagating (when $k_{x}<\omega\sqrt{\varepsilon}/c$,
and $p$ is purely imaginary with $\mathrm{Im}\left[p\right]<0$).
Such condition for signs of the real and imaginary parts of $p$ along
with the positiveness of the sign of the argument of the exponent
$\exp\left[p\left(z+d\right)\right]$ describes the situation, where
the evanescent wave decays in the direction towards $z\to-\infty$,
while the propagating wave propagates in the negative direction of
$z$-axis. 

In other semiinfinite domain $z>d$ the solution of Maxwell equations
(\ref{eq:Max-Ex})--(\ref{eq:Max-Hy}) can be expressed in the form
\begin{eqnarray}
H_{y}^{\left(+\right)}\left(z\right)=H_{y}^{\left(+\right)}\left(d\right)\exp\left[-p\left(z-d\right)\right],\label{eq:Hy-plus}\\
E_{x}^{\left(+\right)}\left(z\right)=-\frac{cp}{i\omega\varepsilon}H_{y}^{\left(+\right)}\left(d\right)\exp\left[-p\left(z-d\right)\right],\label{eq:Ex-plus}\\
E_{z}^{\left(+\right)}\left(z\right)=-\frac{ck_{x}}{\omega\varepsilon}H_{y}^{\left(+\right)}\left(d\right)\exp\left[-p\left(z-d\right)\right],\label{eq:Ez-plus}
\end{eqnarray}
where $H_{y}^{\left(+\right)}\left(d\right)$ is value of the magnetic
field at $z=d$. Owing to the negativeness of the sign of the exponent's
argument $\exp\left[-p\left(z-d\right)\right]$ the wave either decays
towards $z\to\infty$, or propagates in the positive direction of
the axis $z$. 

In the dielectric with losses the solutions of Maxwell equation will
have form
\begin{eqnarray}
H_{y}^{\left(l\right)}\left(z\right)&=&H_{y}^{\left(l\right)}\left(-d\right)\cos\left[k_{z}^{(l)}\left(z+d\right)\right]\label{eq:Hy-d}\\
&&+\frac{i\omega\varepsilon^{(l)}}{ck_{z}^{(l)}}E_{x}^{\left(l\right)}\left(-d\right)\sin\left[k_{z}^{(l)}\left(z+d\right)\right],\nonumber \\
E_{x}^{\left(l\right)}\left(z\right)&=&E_{x}^{\left(l\right)}\left(-d\right)\cos\left[k_{z}^{(l)}\left(z+d\right)\right]\label{eq:Ex-d}\\
&&-\frac{ck_{z}^{(l)}}{i\omega\varepsilon^{(l)}}H_{y}^{\left(l\right)}\left(-d\right)\sin\left[k_{z}^{(l)}\left(z+d\right)\right],\nonumber \\
E_{z}^{\left(l\right)}\left(z\right)&=&-\frac{ck_{x}}{\omega\varepsilon^{(l)}}\left\{ H_{y}^{\left(l\right)}\left(-d\right)\cos\left[k_{z}^{(l)}\left(z+d\right)\right]\right.\label{eq:Ez-d}\\
&&\left.+\frac{i\omega\varepsilon^{(l)}}{ck_{z}^{(l)}}E_{x}^{\left(l\right)}\left(-d\right)\sin\left[k_{z}^{(l)}\left(z+d\right)\right]\right\} ,\nonumber 
\end{eqnarray}
where $k_{z}^{(l)}=\left[\left(\omega/c\right)^{2}\varepsilon^{(l)}-k_{x}^{2}\right]^{1/2}$
, $E_{x}^{\left(l\right)}\left(-d\right)$ and $H_{y}^{\left(l\right)}\left(-d\right)$
being the values of the tangential components of the electric and
magnetic fields at the boundary of the medium with losses $z=-d$. In
the similar manner the electromagnetic field in the medium with gain
can be represented as 
\begin{eqnarray}
H_{y}^{\left(g\right)}\left(z\right)&=&H_{y}^{\left(g\right)}\left(0\right)\cos\left[k_{z}^{(g)}z\right]\label{eq:Hy-g}\\
&&+\frac{i\omega\varepsilon^{(g)}}{ck_{z}^{(g)}}E_{x}^{\left(g\right)}\left(0\right)\sin\left[k_{z}^{(g)}z\right],\nonumber \\
E_{x}^{\left(g\right)}\left(z\right)&=&E_{x}^{\left(g\right)}\left(0\right)\cos\left[k_{z}^{(g)}z\right]\label{eq:Ex-g}\\
&&-\frac{ck_{z}^{(g)}}{i\omega\varepsilon^{(g)}}H_{y}^{\left(g\right)}\left(0\right)\sin\left[k_{z}^{(g)}z\right],\nonumber \\
E_{z}^{\left(g\right)}\left(z\right)&=&-\frac{ck_{x}}{\omega\varepsilon^{(g)}}\left\{ H_{y}^{\left(g\right)}\left(0\right)\cos\left[k_{z}^{(g)}z\right]\right.\label{eq:Ez-g}\\
&&\left.+\frac{i\omega\varepsilon^{(g)}}{ck_{z}^{(g)}}E_{x}^{\left(g\right)}\left(0\right)\sin\left[k_{z}^{(g)}z\right]\right\} ,\nonumber 
\end{eqnarray}
Here $k_{z}^{(g)}=\left[\left(\omega/c\right)^{2}\varepsilon^{(g)}-k_{x}^{2}\right]^{1/2}$,
while $E_{x}^{\left(g\right)}\left(0\right)$ and $H_{y}^{\left(g\right)}\left(0\right)$
stand for the values of the tangential components of the electric
and magnetic fields, correspondingly at the boundary of the medium
with gain $z=0$. 

Boundary conditions for the tangential components of the electric
and magnetic fields can be obtained directly from Maxwell equations
(\ref{eq:Max-Ex})--(\ref{eq:Max-Hy}). Thus, integrating of (\ref{eq:Max-Hy})
over the infinitesimal interval $[d-0,d+0]$ gives the boundary condition
\begin{eqnarray}
E_{x}^{(+)}\left(d\right)=E_{x}^{(g)}\left(d\right),\label{eq:bound-Ex-d}
\end{eqnarray}
which couples the tangential component of the electric field across
the boundary between the gainy and lossless dielectrics. In
the similar manner, integration of (\ref{eq:Max-Hy}) over the intervals
$[-0,+0]$ and $[-d-0,-d+0]$ results in boundary conditions 
\begin{eqnarray}
E_{x}^{(l)}\left(0\right)=E_{x}^{(g)}\left(0\right),\label{eq:bound-Ex-0}\\
E_{x}^{(l)}\left(-d\right)=E_{x}^{(-)}\left(-d\right)\label{eq:bound-Ex-md}
\end{eqnarray}
for the electric field tangential component across the graphene layer
and at the boundary between dissipative and lossless dielectrics, respectively.

Boundary conditions for the tangential component of the magnetic field
can be obtained from integration of Eq.\,(\ref{eq:Max-Ex}) over
the same infinitesimal intervals as in the previous case. The final
expressions can be represented in the form
\begin{eqnarray}
H_{y}^{(+)}\left(d\right)=H_{y}^{(g)}\left(d\right),\label{eq:bound-Hy-d}\\
H_{y}^{(g)}\left(0\right)=H_{y}^{(l)}\left(0\right)-\frac{4\pi}{c}\sigma\left(\omega\right)E_{x}^{(l)}\left(0\right),\label{eq:bound-Hy-0}\\
H_{y}^{(l)}\left(-d\right)=H_{y}^{(-)}\left(-d\right).\label{eq:bound-Hy-md}
\end{eqnarray}
In other words, at boundaries $z=\pm d$ magnetic field is continuous
across the interface, while at the interface $z=0$ the magnetic field
is discontunuous across the graphene due to presence of currents in
it.

Substitution of Eqs.\,(\ref{eq:Hy-minus}), (\ref{eq:Ex-minus}),
(\ref{eq:Hy-d}), and (\ref{eq:Ex-d}) into Eqs.\,(\ref{eq:bound-Ex-md})
and (\ref{eq:bound-Hy-md}) results into
\begin{eqnarray}
H_{y}^{\left(l\right)}\left(-d\right)=H_{y}^{\left(-\right)}\left(-d\right),\\
E_{x}^{\left(l\right)}\left(-d\right)=\frac{cp}{i\omega\varepsilon}H_{y}^{\left(-\right)}\left(-d\right).
\end{eqnarray}
In similar manner, substitution of Eqs.\,(\ref{eq:Hy-plus}), (\ref{eq:Ex-plus}),
(\ref{eq:Hy-g}), and (\ref{eq:Ex-g}) into boundary conditions (\ref{eq:bound-Ex-d})
and (\ref{eq:bound-Hy-d}) gives
\begin{eqnarray}
H_{y}^{\left(+\right)}\left(d\right)=H_{y}^{\left(g\right)}\left(d\right),\\
-\frac{cp}{i\omega\varepsilon}H_{y}^{\left(+\right)}\left(d\right)=E_{x}^{\left(g\right)}\left(d\right).
\end{eqnarray}
In the matrix form the above equations can be represented as
\begin{eqnarray}
\hat{F}^{(l)}\left(-d\right)=\hat{\mathcal{S}}^{(-)}H_{y}^{(-)}\left(-d\right),\label{eq:bound-mat-md}\\
\hat{F}^{(g)}\left(d\right)=\hat{\mathcal{S}}^{(+)}H_{y}^{(+)}\left(d\right),\label{eq:bound-mat-d}
\end{eqnarray}
where $\hat{\mathcal{S}}^{(\pm)}$ and $\hat{F}^{(f)}\left(z\right)$
are 2$\times$1 matrices
\begin{eqnarray}
\hat{\mathcal{S}}^{(\pm)}=\left(\begin{array}{c}
1\\
\mp\frac{cp}{i\omega\varepsilon}
\end{array}\right),\\
\hat{F}^{(f)}\left(z\right)=\left(\begin{array}{c}
H_{y}^{\left(f\right)}\left(z\right)\\
E_{x}^{\left(f\right)}\left(z\right)
\end{array}\right),
\end{eqnarray}
and index $f=g,l$. Along with this, representation of boundary conditions
(\ref{eq:bound-Ex-0}) and (\ref{eq:bound-Hy-0}) in the matrix form
results in 
\begin{eqnarray}
\hat{F}^{(g)}\left(0\right)=\hat{\mathcal{G}}\hat{F}^{(l)}\left(0\right),\label{eq:bound-mat-0}
\end{eqnarray}
where $\hat{\mathcal{G}}$ is the 2$\times$2 matrix 
\begin{eqnarray}
\hat{\mathcal{G}}=\left(\begin{array}{cc}
1 & -\frac{4\pi}{c}\sigma\left(\omega\right)\\
0 & 1
\end{array}\right).
\end{eqnarray}
Along with this, from Eqs.\,(\ref{eq:Hy-d}) and (\ref{eq:Ex-d})
it is possible to link the fields at $z=-d$ and $z=0$ as 
\begin{eqnarray}
\hat{F}^{(l)}\left(0\right)=\hat{\mathcal{T}}^{(l)}\hat{F}^{(l)}\left(-d\right).\label{eq:trans-mat-0md}
\end{eqnarray}
In similar manner from Eqs.\,(\ref{eq:Ex-g}) and (\ref{eq:Hy-g})
one can obtain
\begin{eqnarray}
\hat{F}^{(g)}\left(d\right)=\hat{\mathcal{T}}^{(g)}\hat{F}^{(g)}\left(0\right).\label{eq:trans-mat-d0}
\end{eqnarray}
In above equations the transfer-matrices 
\begin{eqnarray}
\hat{\mathcal{T}}^{(f)}=\left(\begin{array}{cc}
\cos\left[k_{z}^{(f)}d\right] & \frac{i\omega\varepsilon^{(f)}}{ck_{z}^{(f)}}\sin\left[k_{z}^{(f)}d\right]\\
-\frac{ck_{z}^{(f)}}{i\omega\varepsilon^{(f)}}\sin\left[k_{z}^{(f)}d\right] & \cos\left[k_{z}^{(f)}d\right]
\end{array}\right).
\end{eqnarray}

Applying consequently the boundary conditions (\ref{eq:bound-mat-md}),
(\ref{eq:bound-mat-0}), and (\ref{eq:bound-mat-d}) {[}along with
Eqs.\,(\ref{eq:trans-mat-0md}) and (\ref{eq:trans-mat-d0}){]},
we obtain 
\begin{eqnarray}
\hat{\mathcal{S}}^{(+)}H_{y}^{(+)}\left(d\right)=\hat{\mathcal{T}}^{(g)}\hat{\mathcal{G}}\hat{\mathcal{T}}^{(l)}\hat{\mathcal{S}}^{(-)}H_{y}^{(-)}\left(-d\right).
\end{eqnarray}
This equation, being multiplied by row matrix 
\begin{eqnarray}
\left\{ \hat{\mathcal{S}}^{(-)}\right\} ^{-1}=\frac{1}{2}\left(\begin{array}{cc}
1 & \frac{i\omega\varepsilon}{cp}\end{array}\right),
\end{eqnarray}
after taking into account orthogonality of matrices 
\begin{eqnarray}
\left\{ \hat{\mathcal{S}}^{(-)}\right\} ^{-1}\hat{\mathcal{S}}^{(+)}=0
\end{eqnarray}
results in the dispersion relation of the waves in the graphene-based
structure 
\begin{eqnarray}
\left\{ \hat{\mathcal{S}}^{(-)}\right\} ^{-1}\hat{\mathcal{T}}^{(g)}\hat{\mathcal{G}}\hat{\mathcal{T}}^{(l)}\hat{\mathcal{S}}^{(-)}=0,
\end{eqnarray}
which can be written in the explicit form as
\begin{eqnarray}
\frac{\varepsilon^{(g)}}{k_{z}^{(g)}}\Phi^{(g)}+\frac{\varepsilon^{(l)}}{k_{z}^{(l)}}\Phi^{(l)}+\frac{4\pi}{i\omega}\sigma\left(\omega\right)=0,\label{eq:disp-rel}
\end{eqnarray}
where
\begin{eqnarray}
\Phi^{(f)}=\frac{\cos\left[k_{z}^{(f)}d\right]+\frac{\varepsilon^{(f)}p}{\varepsilon k_{z}^{(f)}}\sin\left[k_{z}^{(f)}d\right]}{\sin\left[k_{z}^{(f)}d\right]-\frac{\varepsilon^{(f)}p}{\varepsilon k_{z}^{(f)}}\cos\left[k_{z}^{(f)}d\right]}.
\end{eqnarray}

\section{$\mathcal{PT}$-symmetric surface plasmon-polaritons}

\begin{figure}
	\centering
\includegraphics[width=10cm]{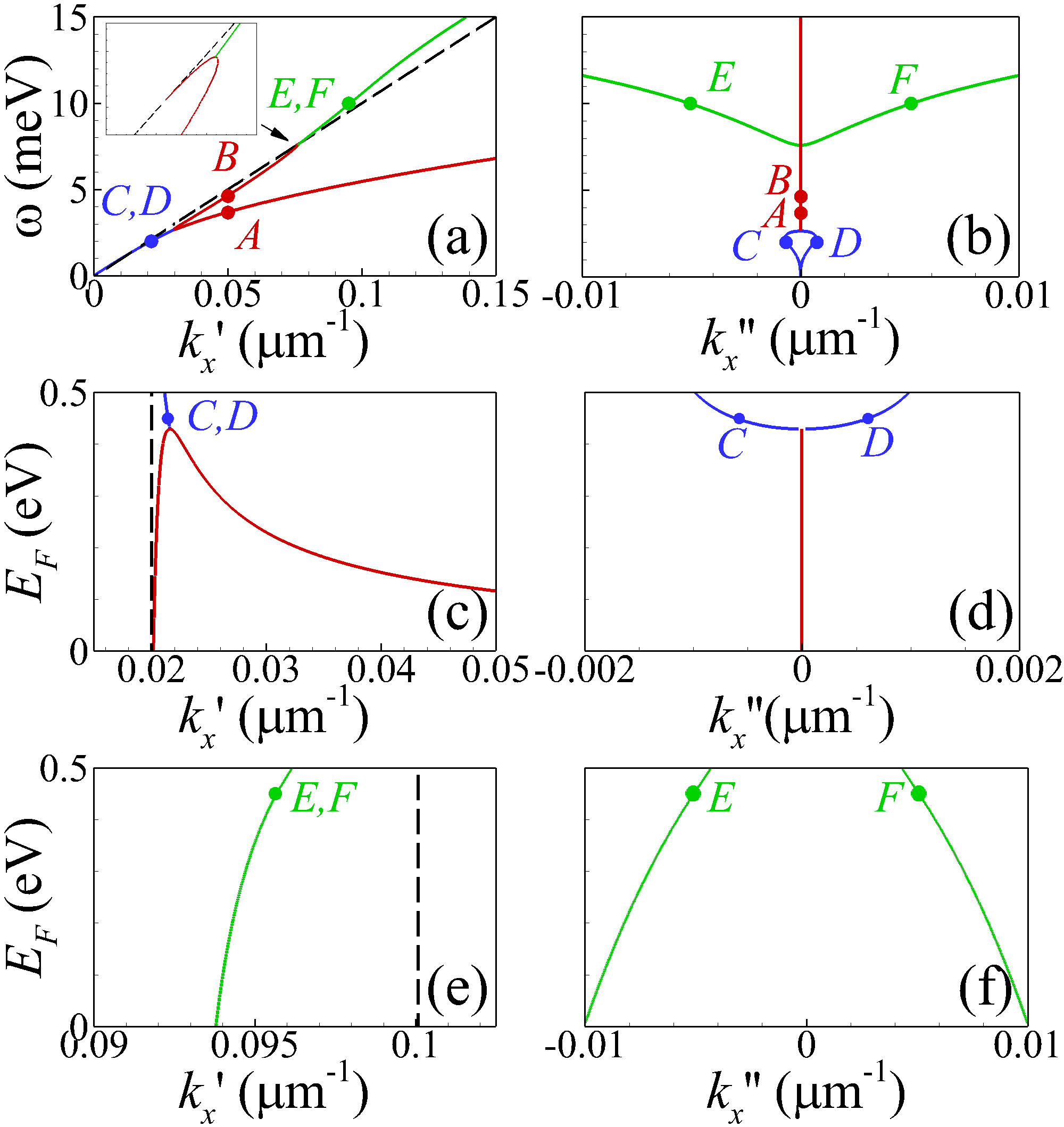}

\caption{(a,b) Dispersion relation of the graphene SPPs in the $\mathcal{PT}$-symmetric
surrounding for fixed Fermi energy $E_{F}=0.45\,$eV -- panels (a)
and (b) present the frequency dependence of the real $k_{x}^{\prime}\left(\omega\right)$
and imaginary $k_{x}^{\prime\prime}\left(\omega\right)$ parts of
the wavevector, correspondingly; (c--f) Dependence of real {[}panels
(c) and (e){]} and imaginary {[}panels (d) and (f){]} parts of wavevector
($k_{x}^{\prime}$ and $k_{x}^{\prime\prime},$respectively) upon
the Fermi energy $E_{F}$ for two frequencies $\omega=2\,$meV {[}panels
(c) and (d){]} and $\omega=10\,$meV {[}panels (e) and (f){]}. In
all panels dispersion curves of graphene SPPs with unbroken $\mathcal{PT}$-symmetry
are depicted by solid red lines, while those with broken PT-symmetry
are depicted by solid blue and green lines. Dashed lines in panels
(a), (c), and (e) stand for the light line $k_{x}^{\prime}=\omega\sqrt{\varepsilon}/c$.
Other parameters of the structure are: $\varepsilon=3.9$, $\varepsilon^{(g)}=\varepsilon^{(l)}=1.9$,
$d=40\,\mu$m, $\gamma=0$.}
\label{fig:sing-lay}
\end{figure}
In the particular situation, when the graphene is considered to be
lossless ($\gamma=0$) and the gain and losses in surrounding media
are prefectly balanced ($\varepsilon_{g}=\varepsilon_{l}$), the dielectric
function (\ref{eq:varepsilon-z}) is characterized by the property
$\varepsilon\left(z\right)=\varepsilon^{*}\left(-z\right)$. In other
words, graphene-based structure possesses the $\mathcal{PT}$-symmetry,
whose distinctive properties are revealed in the dispersion relation
of GSPPs {[}see Figs.\,(\ref{fig:sing-lay})(a) and (\ref{fig:sing-lay})(b){]}.
It should be noted that Figs.\,(\ref{fig:sing-lay})(a) and (\ref{fig:sing-lay})(b)
represent the solution of the dispersion relation (\ref{eq:disp-rel}),
when the frequency $\omega$ is supposed to be purely real value,
while the in-plane wavevector $k_{x}=k_{x}^{\prime}+ik_{x}^{\prime\prime}$
is supposed to be complex value, whose imaginary part characterizes
the degree of exponential decaying (when $k_{x}^{\prime\prime}>0$)
or growing (when $k_{x}^{\prime\prime}<0$) of wave's amplitude per
unit length during the propagation along $x$-axis. 

In the conservative system (without gain/losses) the dispersion relation
of GSPP possesses one branch {[}for details see, e.g. Ref.\,\cite{spp-gr-rev-Bludov2013-ijmfb}{]},
which exists in the whole range of the wavevectors and frequencies.
The situation changes drastically in the case of $\mathcal{PT}$-symmetry.
Thus, in the lossless graphene in $\mathcal{PT}$-symmetric dielectric
surrounding there are two modes of GSPPs with unbroken $\mathcal{PT}$-symmetry
-- low-frequency mode {[}red solid line A in Fig.\,\ref{fig:sing-lay}(a){]}
and high-frequency one {[}red solid line B in Fig.\,\ref{fig:sing-lay}(a){]}.
Unbroken $\mathcal{PT}$-symmetry are characterized by zero imaginary
part of these modes' wavevectors {[}$k_{x}^{\prime\prime}\equiv0$,
see Fig.\,\ref{fig:sing-lay}(b){]}. In other words, these modes
propagate along the graphene layer without damping or growth. At the
exceptional point, located at $\omega\approx2.025\,$meV and $k_{x}^{\prime}\approx0.0218\,\mu\mathrm{m}^{-1}$,
low- and high-frequency modes merge together, and GSPPs with frequencies
below this exceptional point are characterized by broken $\mathcal{PT}$-symmetry.
Here spectrum contains two modes {[}solid blue lines C and D in Figs.\,\ref{fig:sing-lay}(a)
and \ref{fig:sing-lay}(b){]} with complex wavevectors $k_{x}$ such
that for given frequency wavevector of one mode is complex conjugate
of the other mode's wavevector. As a result, one GSPP mode {[}mode
C in Figs.\,\ref{fig:sing-lay}(a) and \ref{fig:sing-lay}(b){]}
is exponentially growing during its propagation along the graphene,
and another mode is exponentially decaying {[}mode D in Figs.\,\ref{fig:sing-lay}(a)
and \ref{fig:sing-lay}(b){]}. At the same time high-frequency GSPP
mode with unbroken $\mathcal{PT}$-symmetry {[}line B in Figs.\,\ref{fig:sing-lay}(a)
and \ref{fig:sing-lay}(b){]} at another exceptional point {[}$\omega\approx7.58\thinspace$eV
and $k_{x}\approx0.0762\thinspace\mu\mathrm{m}^{-1}${]} folds towards
light line $\omega=ck_{x}/\sqrt{\varepsilon}$ and has end-point of
the spectrum lying on this light line {[}which is depicted by dashed
black line in Fig.\,\ref{fig:sing-lay}(a){]}. Along with this, at
that exceptional point {[}see inset in Fig.\,\ref{fig:sing-lay}(a){]}
GSPP mode with unbroken $\mathcal{PT}$-symmetry transforms into pair
of modes with broken $\mathcal{PT}$-symmetry {[}green solid lines
E and F in Figs.\,\ref{fig:sing-lay}(a) and \ref{fig:sing-lay}(b){]},
which degree of growth/decay (imaginary part of the wavevector $k_{x}^{\prime\prime}$)
increases monotonically with an increase of frequency.

One of the advantages of using graphene in plasmonics is the possibility
to tune dynamically graphene's Fermi energy (and, consequently, the
dispersion properties of GSPPs) in time simply by changing the gate voltage applied to graphene. Respective dependence between the Fermi energy and bias voltage $V_b$, applied to graphene, can be expressed as $E_F\sim\left(V_b\right)^{1/2}$ [see, e.g., Refs.\cite{spp-gr-Bludov2012-prb,gr-cond-exp-Li2008-natp,gr-wg-Lu2017-pr}].
In $\mathcal{PT}$-symmetric graphene-based structures it opens another
possibility -- to switch dynamically between the unbroken and broken
$\mathcal{PT}$-symmetries. An example of such situation is demonstrated
in Figs.\,\ref{fig:sing-lay}(c) and \ref{fig:sing-lay}(d), where
for the fixed frequency $\omega=2\thinspace$meV graphene's Fermi
energies above and below $E_{F}\approx0.429\,$eV give rise to the broken
and unbroken $\mathcal{PT}$-symmetries, respectively. Notice, that
upper mode E and F {[}see Figs.\,\ref{fig:sing-lay}(e) and \ref{fig:sing-lay}(f){]}
for chosen fixed frequency $\omega=10\,$meV, and dielectric constant's
imaginary part $\varepsilon_{g}=\varepsilon_{l}=1.9$ exhibits broken
$\mathcal{PT}$-symmetry in all range of nowadays exterimentally attainable \cite{spp-gr-tunable-exp-Ju2011-nn}
graphene's Fermi energies $E_{F}\lesssim0.5\,$eV. Such a mode can
exist even when graphene is absent (case $E_{F}=0$) -- respective
modes were investigated in Ref.\,\cite{pt-waveguide-Kuzmiak2010-oe}.

\begin{figure}
	\centering
\includegraphics[width=10cm]{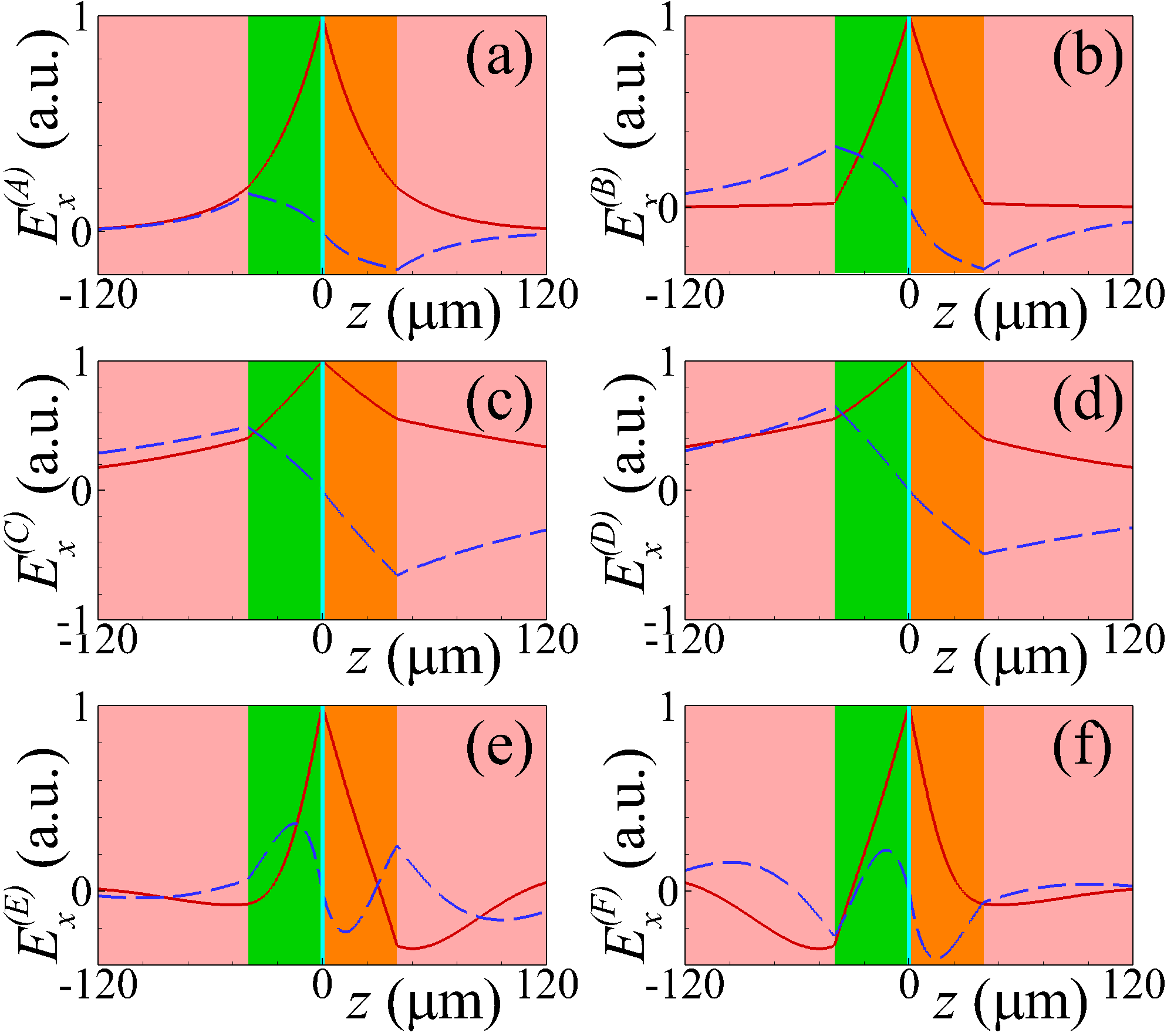}

\caption{Spatial profiles of the electric field in-plane component $E_{x}\left(z\right)$
for frequency $\omega=3.684\,$meV and wavevector $k_{x}=0.05\,\mu\mathrm{m}^{-1}$
{[}panel (a){]}; $\omega=4.642\,$meV, $k_{x}=0.05\,\mu\mathrm{m}^{-1}$
{[}panel (b){]}; $\omega=2\,$meV, $k_{x}=\left(0.0215-i0.0006\right)\,\mu\mathrm{m}^{-1}$
{[}panel (c){]}; $\omega=2\,$meV, $k_{x}=\left(0.0215+i0.0006\right)\,\mu\mathrm{m}^{-1}$
{[}panel (d){]}; $\omega=10\,$meV, $k_{x}=\left(0.0957-i0.0051\right)\,\mu\mathrm{m}^{-1}$
{[}panel (e){]}; $\omega=10\,$meV, $k_{x}=\left(0.0957+i0.0051\right)\,\mu\mathrm{m}^{-1}$
{[}panel (f){]}. Other parameters are the same as those in Fig.\,(\ref{fig:sing-lay})(a).
The frequencies $\omega$ and wavevectors $k_{x}$ of the profles
in panels (a)--(f), are depicted in Fig.\,(\ref{fig:sing-lay})
by points A--F, respectively (which are depicted as superindexes
in y-axis titles). Real and imaginary parts of the electric field
are depicted by solid red and dashed blue lines, respectively. }

\label{fig:sing-lay-fields}
\end{figure}

The physical origins of the reported phenomena can be understood from
spatial distributions of the electromagnetic field, which are shown
in Fig.\,\ref{fig:sing-lay-fields}. In the case of unbroken $\mathcal{PT}$-symmetry
{[}see Figs.\,\ref{fig:sing-lay-fields}(a) and \ref{fig:sing-lay-fields}(b)
for low- and high-frequency modes A and B, correspondingly{]} in-plane
component of the electric field possesses reflective symmetry (real
part, depicted by solid red lines) or antisymmetry (imaginary part,
depicted by dashed blue lines) with respect to the graphene layer,
which is located at the boundary between the gainy and lossy medium
at $z=0$. In other words, their field distribution obey the $\mathcal{PT}$-symmetric
relation $E_{x}^{(A)}\left(z\right)=\left\{ E_{x}^{(A)}\left(-z\right)\right\} ^{*}$,
$E_{x}^{(B)}\left(z\right)=\left\{ E_{x}^{(B)}\left(-z\right)\right\} ^{*}$.
Equality of the field amplitudes in gainy and lossy media originates
a perfect balance between gain and losses of energy during the propagation,
which in its turn lead to the propagation of GSPPs with constant amplitude.
For broken $\mathcal{PT}$-symmetry {[}see Figs.\,\ref{fig:sing-lay-fields}(c)--\ref{fig:sing-lay-fields}(f)
for modes C--F, respectively{]}, the spatial profiles $E_{x}\left(z\right)$
are asymetric. Meanwhile, for modes C and E most of the field is concentrated
in the medium with gain, while for modes D and F most of the field
is concentrated in lossy medium, which determines respective growing
of decaying of energy during mode propagation along the graphene.
At the same time, modes with broken $\mathcal{PT}$-symmetry possess
the mutual symmetry $E_{x}^{(C)}\left(z\right)=\left\{ E_{x}^{(D)}\left(-z\right)\right\} ^{*}$,
$E_{x}^{(E)}\left(z\right)=\left\{ E_{x}^{(F)}\left(-z\right)\right\} ^{*}$,
which determines the equality of absolute values of the imaginary
parts of their wavevectors.

\section{Graphene surface plasmon-polaritons in dielectric medium with unbalanced
gain and losses}

\begin{figure}
	\centering
\includegraphics[width=10cm]{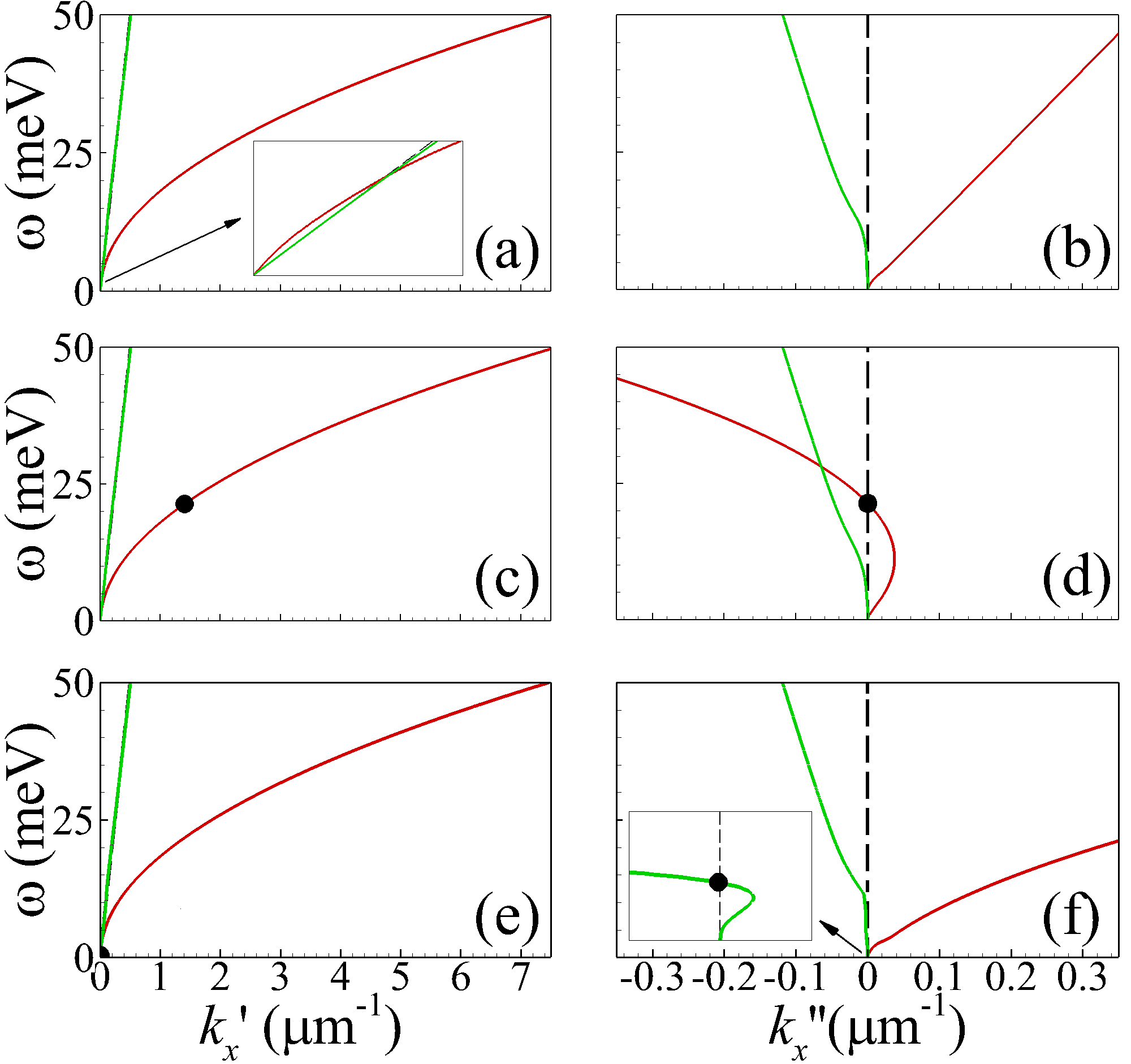}

\caption{Dispersion relation $k_{x}\left(\omega\right)$ of surface plasmon-polaritons,
propagating in lossy graphene monolayer with Fermi energy $E_{F}=0.45\thinspace$eV
and disorder $\gamma=2.5\thinspace$meV, arranged in the dielectric
surrounding with gain/losses $\varepsilon^{(g)}=\varepsilon^{(l)}=1.9$
{[}panels (a) and (b){]}, $\varepsilon^{(g)}=1.9$, $\varepsilon^{(l)}=1$
{[}panels (c) and (d){]}, or $\varepsilon^{(g)}=1.9$, $\varepsilon^{(l)}=3$
{[}panels (e) and (f){]}. Left column {[}panels (a), (c), and (e){]}
corresponds to the real part of the wavevector $k_{x}^{\prime}$,
while the right one {[}panels (b), (d), and (f){]} -- to the imaginary
part $k_{x}^{\prime\prime}$. Other parameters of the structure are
the same as those in Fig.\,\ref{fig:sing-lay}. Black dots in panels (c) and (d) depict the frequency and wavevector of GSPP's spectral singularity.}

\label{fig:sing-lay-with-damping}
\end{figure}
Now a natural question appears: how the dispersion properties will
change, if graphene monolayer is not lossless? To answer this question
we represent in Figs.\,\ref{fig:sing-lay-with-damping}(a) and \ref{fig:sing-lay-with-damping}(b)
the disperion curves of the GSPPs in the case where gain and losses
in surrounding dielectric media are mutually balanced, but small losses
are present in graphene {[}$\gamma\ne0$ in Eq.\,(\ref{eq:drude-cond}){]}.
As evident, losses in graphene results in the situation, where the
GSPP spectrum ceases to be real. In more detail, the low-frequency
mode (solid red lines) is decaying (with positive $k_{x}^{\prime\prime}>0$)
in all range of the frequencies and wavevectors, while high-frequency
one (green solid lines) is growing (with negative $k_{x}^{\prime\prime}<0$).
Nevertheless, if in the dielectric surrounding gain prevails over
losses {[}see Figs.\,\ref{fig:sing-lay-with-damping}(c) and \ref{fig:sing-lay-with-damping}(d){]},
then the low-frequency mode becomes growing at frequencies above some
threshold frequency [$\omega\approx21.38\,$meV for the parameters of Figs.\,\ref{fig:sing-lay-with-damping}(c) and \ref{fig:sing-lay-with-damping}(d)], and remains decaying at frequencies below this threshold. This threshold frequency [called {\it spectral singularity} and depicted in Figs\,\ref{fig:sing-lay-with-damping}(c) and \ref{fig:sing-lay-with-damping}(d) by black dots] plays an important role -- it is the only frequency in all frequency domain, where GSPP spectrum is characterized by purely real wavevector [$k_x\approx1.41\,\mu$m$^{-1}$ for the parameters of Figs.\,\ref{fig:sing-lay-with-damping}(c) and \ref{fig:sing-lay-with-damping}(d)]. In other words, GSPP with frequency, corresponding to spectral singularity, propagates along graphene without growth or damping, owing to its real wavevector. At the same time, when in layered structure losses are higher than gain {[}Figs.\,\ref{fig:sing-lay-with-damping}(e)
and \ref{fig:sing-lay-with-damping}(f){]}, the high-frequency mode 
contains the spectral singularity at frequency $\omega\approx0.44\,$meV. 

\begin{figure}
	\centering
\includegraphics[width=10cm]{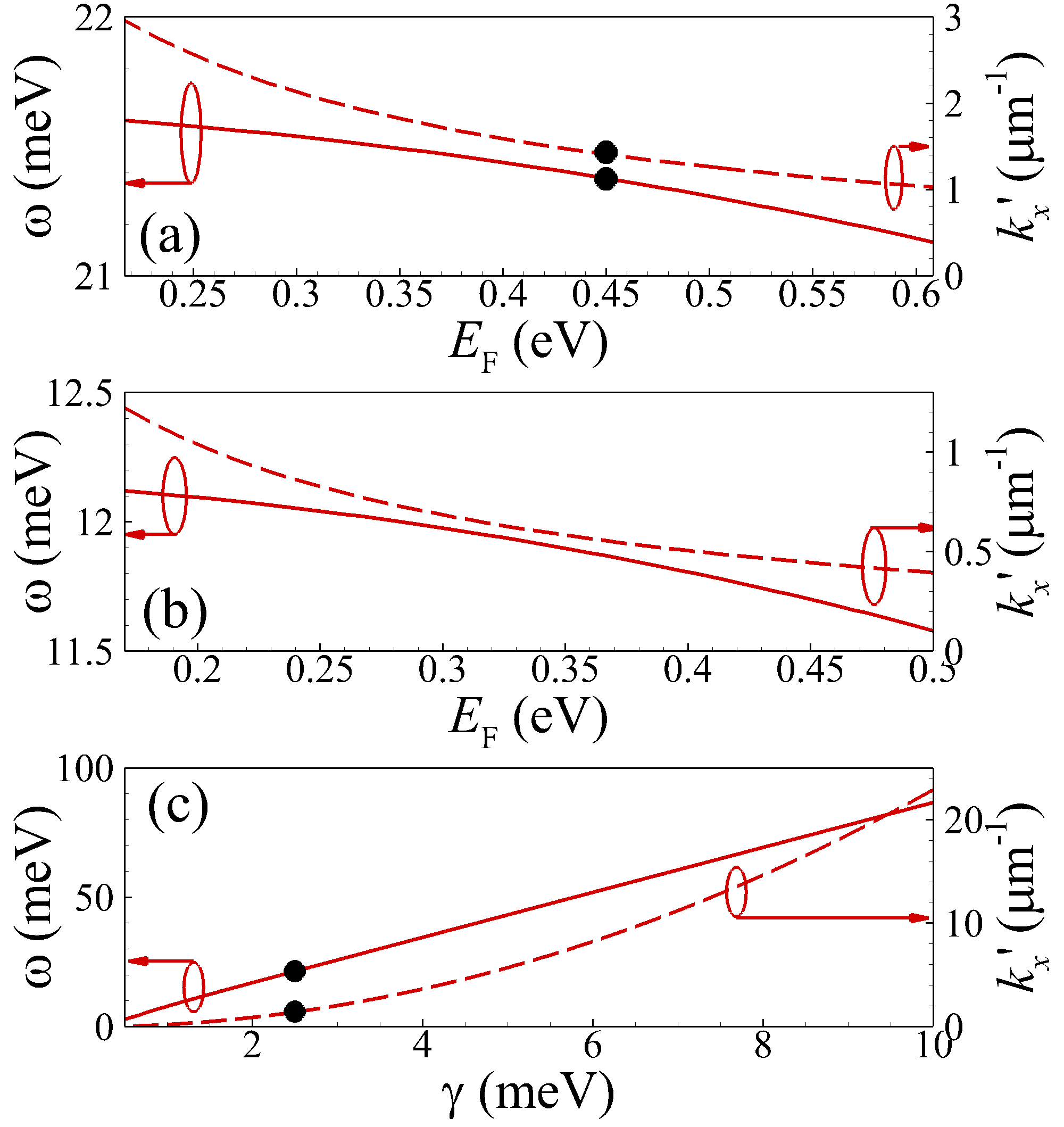}
\caption{Dependence of GSPP spectral singularity's frequency $\omega$ (solid lines, left
axis) and wavevector $k_{x}^{\prime}$ (dashed lines, right axis) upon graphene's
Fermi energy [panels (a) and (b)] or upon the disorder in graphene $\gamma$ [panel (c)]. The parameters of the structure are: $\gamma=2.5\thinspace$meV [panels (a) and (b)], $E_F=0.45\,$eV [panel (c)],
$d=40\,\mu$m, $\varepsilon=3.9$, $\varepsilon^{(g)}=1.9$, $\varepsilon^{(l)}=1.0$
{[}panel (a) and (c){]}, $\varepsilon^{(l)}=0.3$ {[}panel (b){]}. In panels (a) and (c) black dots depict the same parameters of spectral singularity as those in Figs.\,\ref{fig:sing-lay-with-damping}(c) and \ref{fig:sing-lay-with-damping}(d).}
\label{fig:KxEf}
\end{figure}

In connection with this a next question arises: is it possible to
tune the position of spectral singularity 
by varying the Fermi energy? The answer to this question follows directly
from Figs.\,\ref{fig:KxEf}(a) and \ref{fig:KxEf}(b). Thus, for given Fermi energy $E_{F}$ (horizontal axis) 
it is possible to find such frequency $\omega$
(left vertical axis) of spectral singularity, at which GSPP spectrum
will be characterized by purely real wavevector $k_{x}$ (right vertical
axis). Even more, from comparison of Figs.\,\ref{fig:KxEf}(a) and
\ref{fig:KxEf}(b) it follows, that lowering of losses {[}Fig.\,\ref{fig:KxEf}(b){]} in dielectric
leads to the red-shift of the spectral singularity. At the same time for fixed Fermi energy losses in graphene exert strong influence on the position of spectral singularity [see Fig.\,\ref{fig:KxEf}(c)]. Thus, lowering the graphene's disorder $\gamma$ leads to decrease of the respective frequency and wavevector of spectral singularity.

\section{Conclusions}

To conclude, we considered spectrum of GSPPs in the structure, where
graphene monolayer is cladded between two dielectric slabs of finite
thickness -- one slab with gain and another with losses. We demonstrated
that in the case of $\mathcal{PT}$-symmetric dielectric surrounding
the spectrum consists of two modes, which coalescs at exceptional
point. At the frequency ranges below and above the exceptional point
the $\mathcal{PT}$-symmetry is broken and unbroken, respectively.
The position of the exceptional point is sensitive to the Fermi energy
of graphene. Last fact opens the possibility to switch dynamically
between the broken and unbroken $\mathcal{PT}$-symmetry by means
of the electrostatic gating, i.e. by changing the gate voltage, applied
to graphene. When gain and losses in dielectric slabs are not mutually
balanced and graphene is lossy, the GSPP spectrum in such system is
characterized by the presence of spectral singularity -- at certain
particular frequency the GSPP propagation is lossless, i.e. respective GSPP is characterized by infinite mean free path and can travel along graphene monolayer with constant amplitude -- without decaying ou growth. In its turn
the electrostatic gating of graphene (varying the Fermi energy) allows
to change the frequency of spectral singularity. 

\section*{Acknowledgements}

YVB acknowledges support from the European Commission through the
project "Graphene- Driven Revolutions in ICT and Beyond"
(Ref. No. 785219), and the Portuguese Foundation for Science and Technology
(FCT) in the framework of the Strategic Financing UID/FIS/04650/2019.
Additionally, YVB acknowledges financing from FEDER and the portuguese
Foundation for Science and Technology (FCT) through project PTDC/FIS-MAC/28887/2017.

\bibliographystyle{unsrtnat}
\bibliography{pt-single-bib}

\end{document}